\documentclass{emulateapj}
\pdfoutput=1
\usepackage{apjfonts,natbib}

\slugcomment{to be published in Advances in Machine Learning and Data Mining for Astronomy}

\shorttitle{Galaxy Zoo: Morphological Classification and Citizen Science}
\shortauthors{L. Fortson, K. Masters, B. Nichol et al.}

\begin{document}


\title{Galaxy Zoo: Morphological Classification and Citizen Science}


\author{
L.\, Fortson\altaffilmark{1},
K.\,Masters\altaffilmark{2},
R.\, Nichol\altaffilmark{2},
K.\, Borne\altaffilmark{3},
E.\, Edmondson\altaffilmark{2},
C.\, Lintott\altaffilmark{4,5},
J.\, Raddick\altaffilmark{6},
K.\, Schawinski\altaffilmark{7},
J.\, Wallin\altaffilmark{8}
}

\altaffiltext{1}{School of Physics and Astronomy, 116 Church St SE, University of Minnesota, Minneapolis, MN 55455, USA}
\altaffiltext{2}{Institute of Cosmology and Gravitation (ICG), Dennis Sciama Building, Burnaby Road, University of Portsmouth, Portsmouth, PO1 3FX, UK}
\altaffiltext{3}{School of Physics, Astronomy, \& Computational Science, 4400 University Drive, MS 6A2, George Mason University, Fairfax, VA 22030, USA}
\altaffiltext{4}{Oxford Astrophysics, Department of Physics, University of Oxford, Denys Wilkinson Building, Keble Road, Oxford OX1 3RH, UK}
\altaffiltext{5}{Astronomy Department, Adler Planetarium and Astronomy Museum, 1300 South Lake Shore Drive, Chicago, IL 60605, USA}
\altaffiltext{6}{Department of Physics and Astronomy, The Johns Hopkins University, Homewood Campus, Baltimore, MD 21218, USA}
\altaffiltext{7}{Einstein Fellow/Yale Center for Astronomy and Astrophysics, Yale University, PO Box 208121, New Haven, CT 06520, USA}
\altaffiltext{8}{Computational Sciences Program, Department of Physics \& Astronomy,  Campus Box 71, 1301 East Main Street, Middle Tennessee State University, Murfreesboro, TN 37132, USA}



\begin{abstract}

We provide a brief overview of the Galaxy Zoo and Zooniverse projects, including a short discussion of the history of, and motivation for, these projects as well as reviewing the science these innovative internet-based citizen science projects have produced so far. We briefly describe the method of applying en-masse human pattern recognition capabilities to complex data in data-intensive research.  We also provide a discussion of the lessons learned from developing and running these community--based projects including thoughts on future applications of this methodology. This review is intended to give the reader a quick and simple introduction to the Zooniverse.
\end{abstract}



\section{A Brief History of Galaxy Morphology}\label{intro}
\setcounter{footnote}{0}

One of the fundamental facts of the Universe is that most large galaxies\footnote{A galaxy is just a collection of stars; typically billions for a large galaxy.} come in two basic shapes which astronomers call ``Spirals" and ``Ellipticals". The exact details of why this is the case, and how the two types of galaxies relate to each other, remains a major mystery for astronomers. It is central to our understanding of how the creation and evolution of galaxies proceeds with cosmic time and depends on their cosmic location. Significant effort has been spent over the last few decades trying to address these questions.

\begin{figure}
\resizebox{\columnwidth}{!}{\includegraphics{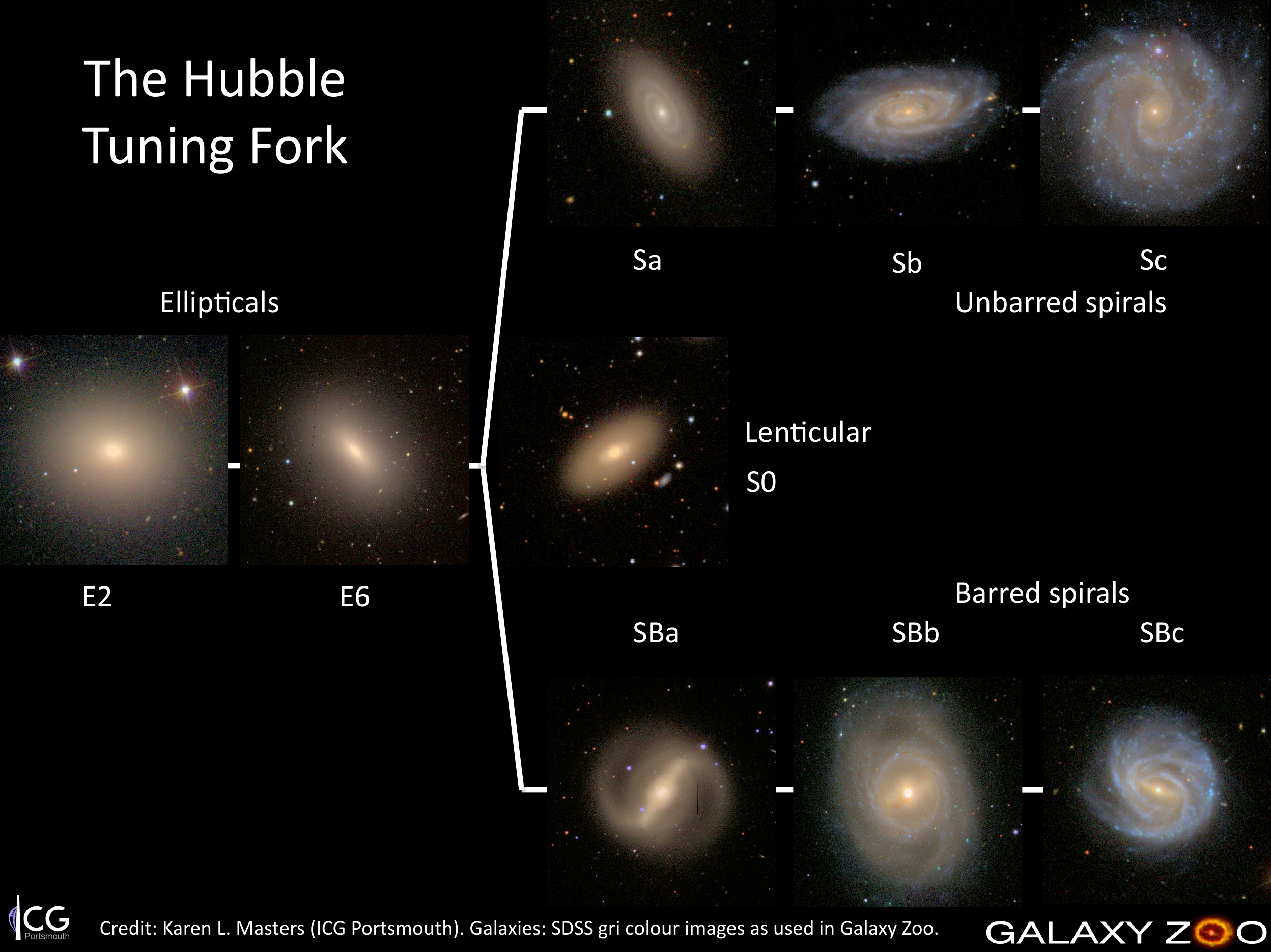}}
\caption{\label{tuningfork} The Hubble Tuning fork illustrated with the type of SDSS colour images used in Galaxy Zoo. Credit: Karen Masters and The Sloan Digital Sky Survey (SDSS) 
Collaboration, www.sdss.org.}
\end{figure}

Edwin Hubble was one of the first astronomers to attempt to systematically address the origin of the shape, or morphology, of galaxies using his famous ``Hubble Sequence" or ``tuning fork" diagram \citep{Hubble1926} which is still in use today (see Figure\ref{tuningfork}). Starting on the left, Hubble classified the elliptical galaxies using the observed ellipticity of the galaxy projected on the sky, giving them a numerical value associated with how round they appeared on the sky. In three dimensions, ellipticals can be triaxial objects, taking a range of morphologies from purely spherical systems through to flattened rugby ball shaped galaxies.

On the right side of the tuning--fork, Hubble placed spiral or disk galaxies. These galaxies have a central ``bulge" of stars, that resemble elliptical galaxies in some ways, embedded in a thin disk of stars that show a range of spiral patterns or ``arms". Hubble ordered disk galaxies based on the tightness of these spiral arms and the size of the central bulge. He had two distinct populations of disk galaxies, namely with and without a central bar-like (or linear) structure. At the point where these different classifications met (for spirals with the largest bulges, and tightest wound arms), Hubble placed ``lenticular" galaxies which at the time were hypothetical disk galaxies with very large bulges and no spiral arms -- they have since been found. 

It is a common misconception that Hubble believed the ``tuning fork" diagram was an evolutionary sequence, with elliptical galaxies on the left evolving along the sequence to form disk galaxies. In fact Hubble advised that ``temporal connotations are made at one's peril" in an early defence of the classification sequence \citep{Hubble1927}, going on to say that he set up the classification ``without prejudice to theories of [galaxy] evolution". This misconception about Hubble's beliefs probably arose due to his suggestion of the use of ``early" and ``late" types to describe the progression towards the right along the sequence (although he discussed that this nomenclature was simply for convenience and borrowed terminology commonly used for stellar classification, \citep{Hubble1926}). Astronomers still call elliptical galaxies ``early" types and disk galaxies ``late" type galaxies, although we now know that most ``late" type galaxies have much younger stellar populations (ironically more ``early type" stars) than most ``early" type galaxies. 

Since Hubble, there have been several updates to his classification scheme \citep[for a recent review see][]{Buta2011} but key features have remained unchanged. What has changed dramatically is the number of galaxies catalogued and requiring classification. Before the advent of digital detectors in astronomy, astronomers could just visually classify the galaxies they saw via their telescopes and/or on photographic plates. New astronomers were trained to follow the classification rules and provided detailed morphologies for thousands of galaxies. Several large catalogues of nearby galaxies with such classifications exists (e.g. The Hubble Atlas of Galaxies \citep{Sandage1961}, or the Third Reference Catalogue of Bright Galaxies (RC3), \citep{deVaucouleurs1991}), and many of these classifications are collected in the NASA/IPAC Extragalactic Database\footnote{\tt http://nedwww.ipac.caltech.edu/}. 

The expert classifier approach quickly became inappropriate with the digital surveys because of the size of the galaxy samples available; for example, the Main Galaxy Sample of the Sloan Digital Sky Survey \citep[SDSS;][]{York2000, Strauss2002} is over a million galaxies and simply cannot be visually inspected by any one astronomer (or even all the astronomers in the world working together). It became clear that some automatic method for classifying galaxies was needed, but programming a computer to recognise the complexities of galaxy shapes (spiral arms, bars, disk plus bulges) is very challenging. 

The first attempts at an automated classification scheme included the use of artificial neural networks, e.g. \citet{Lahav1995} began this work by comparing the classifications of 830 galaxies from a set of six independent experts (R. Buta, H. Corwin, G. de Vaucouleurs, A. Dressler, J. Huchra and S. van den Bergh).  These experts were unanimous in their classification in only 1\% of objects, while an agreement of 80\% between the experts could only be achieved within a spread of two T-types\footnote{A T-type is a numerical coding of the Hubble Sequence increasing from negative numbers for ellipticals and lenticulars, through 0=S0/a, 1=Sa, 3=Sb, 5=Sc and so on.}. The conclusion was that the visual classifications depended on the colour, size and quality of the image used, and that artificial neural networks could be developed to agree to almost the same degree as any pair of expert classifiers. This approach was implemented on the SDSS sample by \citet{Ball2004} with the same basic conclusion as  \citet{Lahav1995}, i.e., that a neural network could reproduce visual morphologies within about 1 T-type.

An alternative approach to developing methods to replicate human classification is to design computational algorithms that attempt to capture the same information. Examples of this approach include the CAS (concentration, asymmetry, clumpiness) structural system of \citet{Conselice2006}, which uses a principal-component analysis (PCA) to study the diversity of internal structures of a sample of galaxies, and the ZEST (Zurich Estimator of Structural Types) algorithm of \citet{Scarlata2007}, which uses a combination of diagnostics of the galaxy shape (that can be measured directly from the galaxy images) and the more traditional Sersic index from the fit to the two-dimensional surface brightness distribution of the galaxy. Such data-orientated methods are very successful at capturing the complexity of galaxy shapes in two-dimensional images but remain hard to translate in terms of the more traditional, established morphological classes discussed above. 

In recent years, there has been significant interest in the development of model--based morphologies of galaxies that use established parametric models for the light distribution of galaxies to fit to the two--dimensional images of galaxies. Such methods include GIM2D \citep{Simard2002} and Galfit \citep{Peng2002} which both attempt to fit galaxy images with a combination of a disk and bulge model. These can be used to construct an objective classification scheme such as the bulge-to-disk ratio of galaxies. Unfortunately, these model-fitting techniques are computationally intense and subject to local minima as they search the high--dimensional parameter space for the best fitting model (in some cases, it can be a 12-parameter model being fit to the galaxy images).  

A quicker way to solve the classification problem is to use a proxy for the galaxy class. The most common such proxy is the colour of a galaxy as most ellipticals are ``red", with their light dominated by older stars, while most spirals are ``blue", as they contain areas of active star formation which include luminous blue stars. However, relying on galaxy classification via colours as a proxy misses an important piece of the galaxy evolution story. The colour of a galaxy is driven by the stellar (and gas and dust) content of the galaxy, while the shape or morphology of a galaxy reflects its dynamical history which could be very different (and have a different timescale). Therefore, one of the central motivations for the original Galaxy Zoo project was to construct a large sample of early and late type galaxy classifications that were independent of colour.

\section{Genesis of Galaxy Zoo}\label{genesis}

The Galaxy Zoo project was inspired by discussions of the limitations of a sample of early-type galaxies produced by \citet{Bernardi2003} from the initial Sloan Digital Sky Survey\footnote{\tt http://www.sdss.org/} data \citep{York2000}. 
\citet{Bernardi2003} had used a PCA-based classification to select ``passive" galaxies based on the spectra of SDSS galaxies. Although this classification scheme was fast, and easy to implement, it probably excluded early-type galaxies that had signatures of on-going star formation. It was therefore realised that to find such objects would require a sample of early--type galaxies based solely on their morphological visual appearance, without the use of spectral or colour information, i.e., that could include normal, passive ``red" early-types, as well as the possibility of ``bluer" star-forming early-types. Kevin Schawinski, as part of his PhD thesis work at Oxford University, under the supervision of Daniel Thomas, took on the task to build such a complete sample of early-type galaxies based solely on their visual appearance and started by inspecting  ~50,000 SDSS galaxies to create the MOrphologically Selected Ellipticals in SDSS (MOSES) sample; an order of magnitude more than any visually inspected sample created to that point. The MOSES sample has resulted in a number of interesting results \citep[e.g.][]{Schawinski2007a, Schawinski2007b, Schawinski2009b, Thomas2010} and in particular shows that there is a significant fraction of early-type galaxies that show recent star-formation activity.  

The experience with MOSES proved the need for independent morphological classifications for galaxies, while also demonstrating that scaling the MOSES methodology to all SDSS galaxies was unfeasible for a small number of researchers to manage. At this point, Kevin Schawinski and Chris Lintott (a researcher at Oxford also involved in MOSES) became motivated to find a way to visually classify all SDSS galaxies in a reasonable amount of time, thus creating the initial ``Galaxy Zoo" concept.
They concluded that the only reasonable way to approach this problem was to ``outsource'' the visual inspection task and put it on the internet
inviting volunteers to participate. At the time, the Stardust@Home Project\footnote{\tt http://stardustathome.ssl.berkeley.edu/} was using the internet to recruit volunteers to identify tracks made by interstellar dust in samples that were 
flown on NASA's Stardust sample-return mission to Comet Wild-2 \citep{Westphal2006}.  Stardust@Home had $\sim20,000$ volunteers, and by extrapolation, Lintott and Schawinski 
figured that if even one quarter of 20,000 volunteers did one galaxy classification per day, the full SDSS Main Galaxy Sample (approximately a million galaxies) could have secure galaxy classifications in three years (assuming each galaxy was visually inspected five times each). 

At the same time, another researcher at Oxford, Kate Land was planning a similar interface to classify, and characterise the sense of rotation of spiral galaxies. She was interested in an article that suggested there was a correlation between the ``handedness" of spiral arms in the SDSS disk galaxies and their position on the sky, i.e., that the direction of the rotation of disk galaxies did not appear to be random \citep{Longo2007}. Land had planned to build an interface on a laptop computer and then place it in the canteen of the Oxford Physics Department, hoping to enlist the help of her fellow scientists. However after a fortuitous meeting of the two groups, it became clear that the projects could be merged into a single interface addressing both questions.

Phil Murray and Dan Andreescu of Fingerprint Digital Media were recruited to design the Galaxy Zoo website and the initial success of Galaxy Zoo can probably be credited to the visual appeal and ease--of--use of the interface design, combined with a relatively easy classification scheme. The user was asked if the galaxy image they saw was ``Spiral" or ``Ellipical", followed by the classification of the apparent spin direction of the spiral arms (clockwise or anticlockwise). Another key factor was that people could get started right away after a relatively short tutorial. Once a user had passed the tutorial, they were free to classify as many galaxies as they wished and could login and out of their account as they wished. The original Galaxy Zoo team along with Land, Lintott and Schawinski included experts from the SDSS (Alex Szalay, Bob Nichol, Steven Bamford, Anze Slozar) and the MOSES team (Daniel Thomas), as well as experts in astronomy outreach (Jordan Raddick) and data archives (Jan van den Berg).

Galaxy Zoo was launched on July 11, 2007 and introduced in a BBC online article that same day\footnote{{\it Scientists seek galaxy hunt help}, by Christine McGourty \\ ({\tt http://news.bbc.co.uk/1/hi/sci/tech/6289474.stm})}.  In the first three hours after launch, classifications were coming in at such a high rate that the data servers located at Johns Hopkins University hosting the site and SDSS images were unable to meet the demand. Fortunately, additional capacity was brought online quickly and, within twelve hours of the launch, the Galaxy Zoo site was receiving 20,000 classification per hour. After forty hours, the classification rate had increased to 60,000 per hour.
After ten days, the public had submitted $\sim8$ million classifications.  By April 2008, when the Galaxy Zoo team submitted their first paper \citep{Lintott2008}, over 100,000 volunteers had classified each of the $\sim900,000$ SDSS galaxy images an average of $38$ times.

One of the unforeseen consequences of the Galaxy Zoo launch was the avalanche of email the team received from the public. Within two weeks, the original Galaxy Zoo team was swamped with requests for information and queries, and several additional people were recruited to help manage these requests. This need to communicate inspired the creation of a Galaxy Zoo internet forum which encouraged the Galaxy Zoo users to communicate with each other (overseen by the Galaxy Zoo team). This allowed many of the basic queries from the public to be answered by other members of the public more experienced with Galaxy Zoo, and also allowed the volunteers (who named themselves ``Zooites") to share their thoughts and ideas with each other.  Once the forum was established, several members of the public (``citizen scientists") quickly volunteered to moderate the forum and began to generate a variety of discussion threads which included basic help with understanding astronomy and Galaxy Zoo, and a repository for ``weird and wonderful" images people found.

In addition to the forum, in December 2007 the team began to communicate with the volunteers through a series of blog messages about the progress of the project and science.\footnote{\tt http://blogs.zooniverse.org/galaxyzoo/}

\section{Galaxy Zoo 1}\label{zoo1}

As described above, the first phase of Galaxy Zoo (now known as ``Galaxy Zoo 1" or GZ1) asked volunteers to provide only basic morphological information on each galaxy. They were asked to identify if a galaxy was ``spiral", ``elliptical", ``a merger" or ``star/don't know" and additionally split the spiral category into ``clockwise", ``anticlockwise" and ``edge-on/don't know". Galaxies for the GZ1 project were drawn from the Main Galaxy Sample of the sixth SDSS Data Release \citep{Strauss2002, Adelman2008} and comprised all extended objects in the survey that were brighter than a Petrosian magnitude of $r<17.77$ mag. All objects were included, whether or not they had an SDSS spectrum, giving a total of 893,212 images.

\subsection{From Clicks to Classifications}

The Galaxy Zoo project was extremely successful in recruiting volunteer classifiers thus providing each galaxy in the sample with multiple independent classifications; GZ1 has a mean of 38 classifications per galaxy, with at least 20 classifications for all galaxy. Most previous morphological classifications had been done by single experts (or small groups of experts) agreeing on a single answer, but in Galaxy Zoo the situation was more like a ``vote" on the galaxy classification. Going from these votes, or ``clicks", to classifications can be done in several ways. 

 The first step in processing the user-generated data was to ``clean" them by removing the tiny fraction of potentially malicious users, and any chance multiple classifications of a single galaxy by a given classifier. Next, there was a decision about how much weight each vote should have. The simplest choice is to give all classifiers equal weight. This gives a distribution of classifications for a galaxy which encodes information about the most likely classification as well as some measure of how certain that is (in the spread of classifications). 

In GZ1 a weighting scheme was also explored which weighted users based on how well they agreed with the majority (in practice this was applied iteratively). This was an attempt to give more weight to ``better" classifiers, where ``better" was defined as agreeing with the majority.
These ``weighted" classifications for the most part were similar to unweighted classifications.

\subsubsection{Classification Biases}

Several bias studies were run in the original GZ1 to test the effect of the interface and types of images shown to the volunteers on the classifications which were entered. The two main goals of the bias studies were to:
(1) test the effect of using colour images for the classifications, and
(2) test if users could reliably identify the sense of the spiral arm winding.
To achieve this, a small number of monochrome and mirrored images were
added to the GZ1 sample and the clicks on these images were compared to
the original, unperturbed images.

Interestingly, a change in the behaviour of the volunteer classifiers was
witnessed during these bias testing exercises, in the sense that users appeared to be more careful in their classifications during bias testing periods. Therefore, only clicks collected on the original images at the same time as the tests were being carried out could be used for the comparison between classifications. The results of the monochrome bias test showed that there were only small differences in the galaxy classifications between colour and black--and--white
images. Users were slightly more likely to classify objects as
``elliptical" in monochrome images; 56\% of the votes went to ellipticals
in the monochrome images compared to 55\% in the original colour SDSS images. 

The results of the mirror image bias testing are discussed extensively in
\citet{Land2008}. They showed a significant bias in favour of
anti-clockwise direction arms (in both the original and mirrored
images). The interpretation of this bias could be due to
psychological effects (possibly related to the preference for right handedness amongst the population), or possibly site design (it being easier to click the anti-clockwise button for example). However, once this bias was corrected for, the data could still be used (see below).

Finally, another source of bias in the GZ classifications has to do with the distance to the observed galaxies. We expect that at some distance,
features become harder to resolve, and more galaxies will be
classified as ellipticals. This effect was indeed found in the GZ1
sample by \citet{Bamford2009} where a correction was derived as a
function of redshift and galaxy luminosity. The conclusion was that
the GZ1 classifications are reliable, and the bias correction is small,
for redshifts below $z<0.08$, but at higher redshifts there is a strong trend for galaxies to be classified preferentially as elliptical.

\subsubsection{Comparison with Other Classifications}

In \citet{Lintott2008}, the GZ1 classifications were compared against three sets of independent galaxy classifications. These included early--type galaxies in the MOSES sample \citep{Schawinski2007b}, a set of 2275 SDSS galaxies of all galaxy types classified by \citet{Fukugita2007}, and the sample of 2834 visually identified SDSS spiral galaxies from \citet{Longo2007}. In all cases, GZ1 classifications were found to agree remarkably well (better than 90\% of the time in most cases), and the conclusion was that using data from volunteers did not substantially degrade the quality of classifications, while expanding the number of classified galaxies by a large factor and additionally reducing the scope for human error introducing erroneous classifications. 

\subsection{Science Results from Galaxy Zoo 1}

Classifications from GZ1 have been used for a wide range of galaxy evolution studies. A full list of the peer--reviewed papers coming from within the Galaxy Zoo 1 team is provided in Table \ref{gz1papers}. We review some of these science results here and stress that the data from GZ1 is now publicly available \citep{Lintott2011}\footnote{\tt http://www.data.galaxyzoo.org} and being used by several scientists beyond the original GZ1 team. For example, Galaxy Zoo 1 was used to remove late-type contaminants from the study of \citet{TFR2011}, and was compared against a new method for automated classification in \citet{Huertas-Company2011}. Moreover, the Galaxy Zoo 1 classifications have now been included in the Eighth Data Release of the SDSS  \citep[see][] {Aihara2011} and can be electronically accessed alongside other SDSS galaxy parameters in their Catalog Archive Server (CAS).\footnote{\tt http://skyservice.pha.jhu.edu/casjobs/}

\subsubsection{Colour and Morphology}
The greatest legacy from GZ1 has been the decoupling of colour
and morphology with high statistical significance. We have demonstrated that 80\% of galaxies follow the expected correlations between colour and morphology, i.e., 
either ``red" early-type galaxies or ``blue" spiral galaxies. Therefore, for a majority of galaxies, colour can be used as a crude proxy for morphology. However, GZ1 also shows that there is a significant numbers of red (passive) spiral galaxies and blue early--type galaxies. These interesting sub-populations of galaxies have been explored in a number of GZ papers (see Table \ref{gz1papers}).

This disentangling of morphology and colour has been used to study the separate dependences of the properties on environment and provide evidence that the transformation of
galaxies from ``blue" to ``red" proceeds faster than the transformation from spiral to early--type (see \citet{Bamford2009} and \citet{Skibba2009} which use different methods to quantify this effect). The properties of the ``blue" early--type galaxies in Galaxy Zoo have been studied by \citet{Schawinski2009a} and ``red" (passive) spirals has been explored further by \citet{Masters2010a,Masters2010b}.

\subsubsection{Spiral Arm Directions}
The clockwise/anti-clockwise classifications of the spiral galaxies
have been used to show that (as expected from the cosmological principle) there is no evidence for a
preferred rotation direction in the universe, but that humans
preferentially classify spiral galaxies as anti-clockwise \citep{Land2008}; and hint at a local correlation of galaxy spins at distances
less than $\sim 0.5$ Mpc - the first experimental evidence for chiral
correlation of spins \citep{Slosar2009}. Intriguingly there are
also hints of a correlation between star formation history and spin
alignments \citep{Jimenez2010}. 

\subsubsection{Merging Galaxies}
The sample of merging galaxies has been used to show that the local
fraction of mergers is about 1-3\% and to study the global properties
of merging galaxies \citep{Darg2010a, Darg2010b}. Multi-mergers (where more than two galaxies are merging at once) - which are much rarer than binary mergers have also recently been studied \citep{Darg2011}. 

\subsubsection{Active Galaxies}
The GZ1 classifications also revealed interesting correlations between galaxy morphology and black hole growth. By splitting both the normal galaxy population and the active galaxy population by morphology, two fundamentally different modes of black hole feeding and feedback in early- and late-type galaxies were found \citep{Schawinski2010b}. Early-type active galactic nucleus (AGN) host galaxies are systematically lower mass and bluer than the general early-type population. Black hole growth is concentrated strongly in the ``green valley" between the blue cloud and the low-mass end of the red sequence. These early-type AGN host galaxies furthermore feature strong post-starburst stellar populations \citep{Schawinski2007b} and thus are migrating from the blue cloud to passive evolution at the low mass end of the red sequence - they are thus building up the red sequence today. 

Late-type AGN host galaxies dominate by number (up to 90\% if "indeterminate" are included) and reside predominantly in massive host galaxies with no indications of recent suppression of star formation. Black hole growth in these disk-dominated galaxies is likely stochastic and has no significant connection to the evolutionary trajectory of the host galaxy. Intriguingly, the Milky Way galaxy resides in the locus of mass and colour where black hole growth is most likely, potentially making the Milky Way and Sagittarius A* a prototype for this ``secular" mode of black hole feeding in late-type galaxies.

\subsubsection{Rare and Unusual Objects}
GZ1 has brought to light several rare classes of object. ``Hanny's
Voorwerp'' is perhaps the most famous of such objects and many are
familiar with the story of the Dutch school teacher Hanny, who first
noted this object (she was not the first volunteer to see it, but the first to ask about it) which is now memorialized in a Comic Book\footnote{see {\tt http://hannysvoorwerp.zooniverse.org/}}. The Voorwerp is an unusual emission line nebula
neighbouring the spiral galaxy IC 2497 and has been studied in several
follow-up projects \citep[e.g.][]{Lintott2009, Rampadarath2010, Schawinski2010a}, and also features in much of the education material from Galaxy Zoo. 

Another unusual class of objects discovered by the Galaxy Zoo volunteers are the ``Green Peas". The properties of these emission--line galaxies, which appear green in the SDSS composite $gri$ colour images because of their strong [OIII] emission, are studied in detail in \citet{Cardamone2009}. 

\begin{table}
\caption{\label{gz1papers} Peer reviewed papers based on classifications collected in the first phase of Galaxy Zoo (in order of publication).}
{\scriptsize \begin{tabular}{ll}
Author \& Year & Title - Galaxy Zoo:  \\
\hline
 Kate Land et al. 2008  & The large-scale spin statistics of spiral galaxies \\ & in the Sloan Digital Sky Survey \\
Chris Lintott et al.  2008  & Morphologies derived from visual inspection of \\&  galaxies from the SDSS \\
Anze Slosar et al.  2009  & Chiral correlation function of galaxy spins \\
Steven Bamford et al. 2009 & The dependence of morphology and colour \\&on environment \\
Kevin Schawinski et al. 2009a & A sample of blue early-type galaxies at low redshift \\
Chris Lintott et al. 2009  & `Hanny's Voorwerp', a quasar light echo? \\
Ramin Skibba et al. 2009  & Disentangling the environmental dependence of \\&morphology and colour \\
Carie Cardamone et al.  2009  & Green Peas: discovery of a class of compact \\&extremely star-forming galaxies \\
Danny Darg et al.  2010a  & The fraction of merging galaxies in the SDSS and \\&their morphologies \\
Danny Darg et al. 2010b  & The properties of merging galaxies in the nearby \\& Universe - local environments, colours, masses, \\& star formation rates and AGN activity \\
Kevin Schawinski et al. 2010a  & The Sudden Death of the Nearest Quasar \\
Kevin Schawinski et al. 2010b  & The Fundamentally Different Co-Evolution of \\& Supermassive Black Holes and Their Early- and \\& Late-Type Host Galaxies\\
Karen Masters et al.  2010a  & Dust in spiral galaxies\\
Raul Jimenez et al. 2010  & A correlation between the coherence of galaxy spin \\& chirality and star formation efficiency \\
Karen Masters et al. 2010b  & Passive red spirals \\
Manda Banerji et al,  2010 & Reproducing galaxy morphologies via machine learning \\
Chris Lintott et al 2011  & Data Release of Morphological Classifications for \\& nearly 900,000 galaxies \\
O. Ivy Wong et al. 2011  & Building the low-mass end of the red sequence with \\& local post-starburst galaxies \\
Daniel Darg et al. 2011 &  Multi-Mergers and the Millennium Simulation
\end{tabular}}
\end{table}

\section{Evolution of Galaxy Zoo}\label{evolution}

\subsection{Galaxy Zoo 2 and Hubble Zoo}

As the original Galaxy Zoo was the first time such a project had been attempted, the Galaxy Zoo team was cautious with their classification scheme, only asking for simple information about the appearance of the galaxies. Thanks to the overwhelming response, and prompted by requests from the volunteers who wanted to provide more detailed classifications,  the team realized they could harvest much more information from the SDSS images than in GZ1. Therefore, Galaxy Zoo 2 (GZ2) was designed around asking more detailed questions about the $\sim250,000$ brightest SDSS galaxies from the original GZ1 sample of galaxies. \footnote{The website ({\tt http://zoo2.galaxyzoo.org/}) for this phase of Galaxy Zoo was designed by Phil Murray and implemented by Danny Locksmith and Arfon Smith.} Once again, the response was tremendous 
and in the fourteen  months the site was live, Galaxy Zoo 2 users provided over 60 million classifications. Along the way, deeper SDSS images were added for a subset of GZ2 galaxies, taken from a patch of the sky known as ``Stripe 82" which allows fainter structures in these galaxies to be visible. 

The first science results from GZ2 classifications are now appearing. In \citet{Masters2011}, we showed that the fraction of barred disk galaxies (as compared to unbarred galaxies) depends on other galaxy properties, especially the overall colour of the galaxy and the size of the central bulge. As a satellite project, Ben Hoyle at Portsmouth University developed an additional web interface using Google Maps technologies to allow GZ2 volunteers to draw the shapes and sizes of bars on GZ2-selected disk galaxies 
\citep{Hoyle2011}. From September 2009 to January 2010, he received 16,551 bar drawings for 8180 galaxies, making it by far the largest sample of disk galaxies with known bar lengths; again demonstrating the attraction of Galaxy Zoo even for such a complex task. These studies combined show the strong connection between the bar of a disk galaxy and its overall colour, i.e., disk galaxies with long bars also exhibit prominent bulges and have redder colours than galaxies with smaller bars.

After Galaxy Zoo 2, the team launched ``Hubble Zoo". To really understand galaxy evolution, and to get a
sense of how the colour-morphology relation might change over time, it is important to be able to classify morphologies for galaxies that
are much further away than those classified from the SDSS. The light from these galaxies has taken much longer to get to us and hence provide images of galaxies at a much earlier epoch
in the history of the universe. Such a dataset will allow us to answer questions like: Are there more blue ellipticals compared to red ellipticals earlier on in the Universe? Does the number of irregularly shaped galaxies increase as we look back further in time? To compare the results from the GZ1 and GZ2 classifications of the SDSS galaxies to galaxies at an earlier epoch,
the latest incarnation of Galaxy Zoo is using data from the Hubble Space Telescope (HST) which goes deeper than ever before, e.g., HST COSMOS (Cosmic Evolution Survey) has over two million galaxies
that cover 75\% of the age of the universe \citep{Scoville2007}. Hubble Zoo is currently undergoing classification using HST data from GEMS \citep{Rix2004}, 
GOODS \citep{Giavalisco2004}, AEGIS \citep{Davis2007} and COSMOS \citep{Scoville2007}.  The decision tree is identical to that for GZ2 except that there is an additional branch that classifies the ``clumpiness" and symmetry of each galaxy.

\subsection{The Citizen Scientists - Motivation and Unexpected Outcomes}
Within the first several days after launch, it was clear to the Galaxy Zoo team that they had hit a nerve with the public - classifying galaxies on Galaxy Zoo provided some sort of 
fulfilment for the volunteers. GZ team members suspected that the popularity of the project relied on the beauty of the images, or that the project
had benefited from particularly good and lucky publicity. Already, team members were thinking of other scholarly areas where applying the method of visually inspecting 
data could lead to publishable results beyond what could be accomplished by application of machine algorithms. But before any such steps could be taken, it was essential
to understand the motivations for volunteers participating in Galaxy Zoo.   A survey of the motivations of citizen scientists involved in Galaxy Zoo is presented in 
\citet{Raddick2010}.  The results show that by far the most common motivation Galaxy Zoo volunteers cite for their involvement in the project is their desire to contribute to real scientific work. 

Thus it should not have come as a surprise that many Galaxy Zoo volunteers developed their own lines of inquiry off the main task page. 
The Galaxy Zoo forum acted as a clearing house for volunteers to describe and discuss objects that they felt were noteworthy. Several threads were devoted to collecting objects with specific characteristics, e.g. triple mergers,
 or  ``overlapping" galaxies, or small, round, green galaxies dubbed ``Peas". These three examples have all resulted in scientific papers \citep[respectively]{Darg2011, Keel2011, Cardamone2009}. 

But one of the critical aspects enabling the development of collections and further inquiry into an object's characteristics was the link from
the main task page for each object to the SDSS SkyServer Object Explorer page\footnote{\tt http://skyserver.sdss.org/dr8/en/tools/explore/obj.asp}. This page aggregated information about the galaxy including the
image and accompanying spectrum as well as information
about its magnitude, redshift, cross-identifications in other wavelengths and a host of links to more information such as NASA's Extragalactic Database\footnote{\tt http://nedwww.ipac.caltech.edu/}.
It is through the Object Explorer page that Galaxy Zoo volunteers began to notice that the ``Peas" all had extraordinarily high fluxes in the [OIII] emission line. 
Eventually over 250 of these objects were found while volunteers taught each other through the forum what the characteristics of a ``pea" were and
began to trade literature searches on what [OIII] meant and possible interpretations of these galaxies. After several months collecting and interpreting on their own, a graduate
student from Yale, Carie Cardamone was assigned to moderate the ``Peas" forum, working with the volunteers while she developed the full analysis
of these rare dwarf galaxies with extremely high star formation rates \citep[which was published as][]{Cardamone2009}. 

The story of the Galaxy Zoo ``Peas" inspired the team to ensure that future projects provide links to supporting information and analysis tools
related to the objects shown in the primary task. This is to enable the users to conduct their own research  and allows for users to learn the
process of research aided by peer-mentoring.

The experience with the Galaxy Zoo forums and blogs shows that the citizen scientist volunteers wanted to do much more than classify objects. They
built a community of the volunteers, by the volunteers and for the volunteers. Indeed, Galaxy Zoo belonged to the volunteers - it was their time just as
much as it was the scientists time spent working on the project. The team understood this important fact and made it a point of principle to keep the volunteers informed
about various aspects of the project from the technical to the social and scientific. Moreover, the team realised early on that they must respect the time and commitment of the volunteers, and should only harvest classifications for as long as they were scientifically useful. 

In fact the volunteers have set up several projects of their own using Galaxy Zoo infrastructure or methods. The largest example of such a project is probably the ``Irregulars" project. Initiated by Galaxy Zoo volunteers Richard Proctor (``Waveney") and Julia Wilkinson (``Jules") on a forum thread\footnote{\tt http://www.galaxyzooforum.org/index.php?topic=273410.0}, the aim of this project was initially to collect a sample of irregular galaxies, i.e. galaxies that did not fit in to the classification scheme at all. This project now uses a self built web interface (similar in style to Galaxy Zoo)\footnote{\tt http://www.wavwebs.com/GZ/Irregular/Hunt.cgi} to ask for classifications of the objects and has inspired several volunteer led research papers. Richard Proctor has recently applied to do a part-time PhD at the Open University using the data collected in this project. 

 The Galaxy Zoo forum\footnote{\tt http://www.galaxyzooforum.org} has been a scientific gold mine on several occasions. Examples of science results coming directly from the forum include the discovery of the Voorwerp \citep{Lintott2009}, targeted and serendipitous searches for smaller versions of the Voorwerp \citep[``voorwerpjes"][]{ChojKeel2011, Gagne2011}, the Peas (Cardamone et al. 2009), overlaps (Keel et al. 2011), ring galaxies, etc. The depth of interest shown by some of the volunteers is extraordinary. Volunteer, Richard Proctor (``waveney") has set up web forms for several searches and sample evaluations (including the ``Irrregulars" project mentioned above). Massimo Mezzoprete (``Half65") was so interested in the overlapping-galaxy search (Keel et al. 2011) that he learned SQL and perl, creating a tool that he could point to a forum thread and have it parse for either kind of unique SDSS Object ID, then query the Catalog Archive Server and create a PDF with a page of finding chart, photometry and positional data for each object. (Mezzoprete is a co-author on the first overlapping-galaxy paper). These forum results clearly show that through the Galaxy Zoo project, citizen scientists have become research collaborators.

\section{The Zooniverse}\label{zooniverse}

The extension of the Galaxy Zoo idea to other scientific domains is obvious in our data-rich world, especially given the desire
of the public to be involved in scientific investigations of these data.
Researchers across a diverse range of academic fields face the common problem of developing new strategies and modes of computational thinking
needed to transform this data flood into knowledge. With the current moderate-sized databases (terabytes), citizen science methods like Galaxy Zoo can replace some aspects
of machine algorithms. However, as the data deluge will only intensify in the next decades, machine algorithms must advance to meet the data processing demands,
incorporating techniques based on developing areas such as computer vision. Instead of displacing the citizen science method, these new algorithms will
need to be trained from, and tested by, human input (e.g. GZ1 classifications have been used for machine learning in \citet{Banerji2010}). Thus, the visual processing methods of Galaxy Zoo will become essential to fully extract information from the data.

It is tempting to think of Galaxy Zoo purely as an Education and Outreach endeavor with all its successes
in garnering publicity and focus on a community of non-expert volunteers. And with that temptation, one
might imagine applying the Galaxy Zoo method to an indiscriminate array of projects with the idea that
the public would be engaged in the process so it does not matter if the scientific outputs were ``real" or whether
the data processing could have been better accomplished through standard computational methods. What must be
made clear is that Galaxy Zoo turned citizen science into a data processing \emph{method} - a
data reduction tool for data-intensive science
which when applied correctly provides the best possible data product from a set of ``raw" data. The genius in
this method lies in the fact that the public actually prefer to participate in a meaningful set of tasks where they
know their work is useful.  Galaxy Zoo established this coupling between high-priority science output and the public
engagement in science. Once it became clear that the appetite of the volunteer classifiers could crunch significantly more
data the question became one of how this new citizen science method could be made available across different disciplines and data products.
And how to begin the process of developing the machine algorithms trained by the human classifiers.

The first step in providing access to more, and varied, data-intensive projects was to aggregate individual citizen science projects onto a common web-based
portal. Several objectives are met by
establishing a centralized common entry point. First, a home base is provided for the volunteers so they can move with ease between projects. It encourages a sense of
community as the same volunteers can share information about their work within, and across, projects. It builds confidence and ``brand loyalty" allowing volunteers to become
more willing to try new types of projects and progress in their learning. Second, aggregating projects
allows for cyberinfrastructure that can take advantage of cross-project efficiencies while retaining the flexibility to provide individualized tools for specific projects.
For example, shared software makes development of different projects possible on a reasonably short timescale thus reacting quicker to new opportunities. Among other advantages provided, these factors
can then reduce the overhead on recruiting volunteers and allows for the possibility of deploying
small and exploratory projects that would be prohibitive to create on their own.   Building on the zoo aspect of the Galaxy Zoo brand, the  ``Zooniverse"
became the answer to how to create a centralized portal to a universe of Zoo-like projects.

To turn the ``Zooniverse" into reality, several new projects with data sets beyond the SDSS were developed. 
In order to help manage the Zooniverse and its expanding set of projects, in June 2009 the Citizen Science Alliance\footnote{see {\tt http://www.citizensciencealliance.org}}  was formed initially by Chris Lintott, Steven Bamford, Lucy Fortson and Arfon Smith. The Zooniverse Project\footnote{\tt http://www.zooniverse.org} website was launched in December 2009.

To shift from the original Galaxy Zoo to the Zooniverse, substantial technical changes were implemented in order to produce a robust and flexible system. 
The most important change was the shift from hosting on a single server to hosting in the ``Cloud", i.e., making use of commercial services provided by Amazon Web Services. 
This technology allows new servers to be brought online in response to demand, and therefore allows the site to cope with spikes in internet traffic due to the fluctuating media coverage. 
The new system is built in a ``Ruby on Rails" framework with a restful API layer between a thin web layer and the database. Authentication of users is carried out by an 
implementation of the Central Authentication Service (CAS) single sign-on solution. This technology allows volunteers to use the same account for both the forum and the main Galaxy Zoo site, as 
well as between different projects. The use of an API allows the Zooniverse team  to support not only the main website but also iPhone and Android applications, allowing mobile 
users to take 
part in Galaxy Zoo. Early results suggest that this may be an effective way of increasing the number of classifications per user.

The Zooniverse codebase was designed with a flexible domain model and extensible reuse of code. These attributes allow
features developed for new projects to be useable by all projects. 
The use of cloud computing services provides hosting scalability, while the virtual platform also handles content distribution and asynchronous classification processing.
As of early 2011, the ``Zooniverse" is running eight Zoo projects and has handled many millions of classifications by more than 250,000 users.
Several different task functions have been implemented through these projects including basic decision trees, drawing
shapes on images (``MoonZoo", ``MilkyWay" Project), real-time asset prioritization and
alerts with the Galaxy Zoo Supernova project of \citet{Smith2011}, manipulating simulated data parameters (Galaxy Zoo Mergers) and text transcription
(``Old Weather").

To aid in the development of the Zooniverse as a community of citizen scientists, and to enable users to engage in inquiry related to the data for a given Zoo, 
a discussion tool was recently developed to replace the forum structure used in GZ.
The new discussion tool (called ``Talk") was launched with the Milky Way Project and encourages users to create collections of objects, share information and 
join in online discussions. Several social media features such as tagging, tag clouds, ``trending" and ``recent" toggles improve Talk over the older forum structure,
while retaining the primary collaborative functions such as the discussion boards in Galaxy Zoo. The Zooniverse team has already seen a marked increase in traffic to Talk compared
to the number of users navigating to the old forum structure. 

The Zooniverse team also has developed numerous education resources and continues to conduct education research into the motivations of the volunteers
to contribute to tasks, how usage patterns vary over different levels of engagement with the project and whether there is any gain in understanding the process
of research - just to name a few of the topics. Further description of these efforts is outside the scope of this paper.

\subsection{Tasks suitable for the Zooniverse}

One of the difficulties for the Zooniverse is
understanding the types of tasks that are suitable for citizen scientists.   The original
Galaxy Zoo project primarily asked users to
classify images.   The interface was simple, with only a few
buttons to click for every image.   Some of the early success of this project 
might have been due to the simple requirements of this task.

The newer Galaxy Zoo 2, Hubble Zoo and Galaxy Zoo Supernova project are also based on having volunteers do classifications on images.
However, these projects use a context-based decision tree to ask
more detailed morphological questions about the objects, rather than just
using a single classification of an object.    If the galaxy
was a spiral, does it have a bar at the center?  How many spiral arms are visible? 
Although each question only has a few possible answers, the 
data for each object has more detail than would be possible from a single
question interface.

The GZ ``Mergers" project operates in a fundamentally different way than
the others Zoos. Users are
not asked to classify images, but rather to match simulated images to data.   In 
some cases, none of the simulations presented are similar to the target galaxy.
In other cases, there are several selections possible within the main interface.
After selecting an image, users then have the option of enhancing it using
a Java applet.   By using two dimensional sliders, the users can generate new simulations
to try to make their results match the target image more closely. The overall approach of this project has some similarities
to on-line citizen science games like 
``Fold-It!".\footnote{\tt http://fold.it/portal/}   The primary difference is the lack of an objective score for the goodness--of--fit.   In the case of mergers, we do not have such an objective fitness function, as one of the goals of the project is to create a sufficiently large sample of galaxy mergers that such a function could be derived.  The users therefore have to use their best judgement to determine the goodness--of--fit.

When the GZ ``Mergers" project first started, there were a large number of images that were
being viewed every day.   Of the images being viewed, approximately 5\% were selected 
by the volunteers as possible matches to the target galaxy.   Upon inspection, the science team
found that a high fraction (up to 95\%) of the simulations were not likely
matches to the real galaxies, as it appeared that many simulations were inadvertently
selected, or inexperienced users tried to select too many simulated galaxies as matches.   
To increase the fraction of good matches, the team created a second
level interface called ``Merger Wars''.    In this interface, volunteers were given
the opportunity to select the best of the simulation images by allowing the full suite of simulated images to compete with each other in one--to--one competitions, e.g., users were shown only two simulations at a time, and asked to pick the best one, and then iterate.  Although some
of this analysis is still underway, the science team believes that the 
selection rate for good matches has dramatically improved.     A larger fraction
of the originally selected simulations get zero votes in this second level competition.

In the Planet Hunters\footnote{\tt http://www.planethunters.org/} site, users are not looking at images, but rather are presented with time series data on the light curves of nearby stars, and are tutored on how to recognize the signature of an extrasolar planet in that data. Despite the seemingly esoteric nature of light curve data this Zooinverse project has been very successful, proving that citizen scientists are happy to deal with more complex types of data possibly because the scope for discovery is high. 

In addition to classification and matching, citizen scientists are also being asked
to do measurements on images. In the ``Solar Storm Watch", ``Moon Zoo", and ``Milky Way" projects,
volunteers use drawing tools to identify features.   In the ``Moon Zoo", for example,
volunteers are asked to draw circles around the rims of craters.   A similar process
is used to identify bubbles in the interstellar medium in the ``Milky Way" project.

In some ways, these last projects require more advanced skills and more patience
than just clicking through a classification tree.   However, with the right interface
and the right users, very good results can be obtained on these types of projects.

A key observation from all of these Zooniverse projects, and from the forums and Talk interactions, is that
some of the volunteers have very advanced abilities and interests.   
There is a great deal of effort being dedicated to develop a suite of tools that allow these users to do additional scientific investigations on their own and, as discussed above, some of the most interesting discoveries come from the users themselves. 

\subsection{Data Mining the Zooniverse Results}
One of the key features of the Zooniverse project is the application of machine learning (data mining) algorithms to the Zooniverse volunteer-contributed tags.  These tag data themselves generate a significant volume of data (e.g., the many hundreds of millions of galaxy classifications from Galaxy Zoo).  Finding correlations and trends among these user-contributed tags alongside automatically measured parameters of the same objects within the science database (e.g., the SDSS object catalog) will enable the development of improved classification and anomaly-detection algorithms for future sky surveys (such as the Large Synoptic Survey Telescope (LSST)), which will measure properties for at least 100 times more galaxies, 100 times more stars, and 100 thousand times more source observations.

For example, a preliminary study of the galaxy mergers found in the Galaxy Zoo I project was carried out \citep{Baehr2010}.  It was found that certain science database parameters in the SDSS science database correlated strongly with how often Galaxy Zoo users identified an object as a merger.  These database attributes included: (a) the log-likelihood that the galaxy's surface brightness profile was fit neither by an exponential disk (the {\tt lnLExp\_u} attribute in the {\tt PhotoObjAll} table)  that is typical of spiral/disk galaxies nor by a de Vaucouleurs profile (the {\tt lnLDeV\_u} attribute in the {\tt PhotoObjAll} table) that is typical of elliptical galaxies; (b) a gradient in the position angle of the isophotal major axis of the galaxy (the {\tt isoAGrad\_u} attribute in the {\tt PhotoObjAll} table up to Data Release 7); and (c) the galaxy's ``texture" (the {\tt texture\_u} attribute in the {\tt PhotoObjAll} table up to Data Release 7), which is essentially the RMS (root-mean-square) variation of the galaxy's surface brightness profile relative to one of the standard galaxy profile-fitting functions.  In hindsight, it could have been predicted that these parameters would be useful in distinguishing normal (undisturbed) galaxies from abnormal (merging, colliding, interacting, disturbed) galaxies.  These results may now be applied to future sky surveys, to improve the automatic (machine-based) classification algorithms for colliding and merging galaxies.  All of this was made possible by the fact that the galaxy classifications provided by Galaxy Zoo I participants led to the creation of the largest pure set of visually identified colliding and merging galaxies yet to be compiled for use by astronomers.

Another example of machine learning using Galaxy Zoo classifications is provided in \citet{Banerji2010} who trained a neural network on a subset of the GZ1 data , and (depending on the automatic measurements given to the algorithm) could reproduce the classifications to better than 90\%. They concluded that Galaxy Zoo would provide an invaluable training set for future algorithms likely to be developed to classify the next generation of wide-field imaging surveys. 

\subsection{Future citizen science projects}

With all the recent activities in the Zooniverse, it is important to consider the implications that citizen science has for future astronomical projects (e.g. LOFAR \citep{Falke2007}, and the Dark Energy Survey\footnote{\tt http://www.darkenergysurvey.org}). For example, we briefly consider here how volunteers might help with a project like the Large Synoptic Survey Telescope (LSST) \citep{LSST2009}, when it comes online this decade.

During the first Galaxy Zoo project, volunteers examined images of approximately one million galaxies. The storage space needed for all of these images was only a few Terabytes and therefore relatively easy to host and serve.  In contrast, LSST will generate tens of Terabytes per day and over its approximate ten year operational lifetime, it is estimated that it will generate tens of Petabytes.

Among the citizen science projects that may contribute to LSST science are those that explore the time series data from the survey.  Since LSST will do repeated imaging of the sky over the 10-year project duration, each of the roughly 50 billion objects observed by LSST will have approximately 1000 separate observations.  These 50 trillion time series data points will provide an enormous opportunity to discover all types of rare phenomena, rare objects, rare classes, and new objects, classes, and sub-classes. 

No group of volunteers could hope to view all such data being generated from LSST. At the same time, the science team on the project will have no hope to keep up with such a data flow from the system.    Obviously, automatic algorithms need to be used to triage
the data and do basic classification of events.   However, even with automatic classification, 
it is anticipated that tens of thousands (or more) anomalous events will be detected every day.   
Some of these might be astronomically significant (asteroids, supernovae, etc).   However,
many will not fall into any particular category, and in many cases, they might be some kind
of noise (an airplane flying in the field of view).   

The contributions of human participants may include: detection of unusual light curves in rotating asteroids; human-assisted search for best-fit models of these asteroids (including shapes, spin periods, and varying surface reflection properties); discovery of unusual variations in known variable stars; discovery of interesting objects in the environments around variable objects; discovery of associations among multiple variable and/or moving objects in a field; and more.  This is especially important for the nightly event stream -- perhaps 100,000 new events will be detected each and every night for 10 years.  There are not enough observing facilities or professional astronomers (or graduate students) in the world to follow up on each of these events.  Engaging a large cohort of willing participants to examine these events will contribute significantly to the scientific discovery efficiency and effectiveness of the LSST survey:  citizen scientists may explore this massive event stream for novel and interesting features, thereby characterizing the behavior of each such object.  The creation of a ``characterization database'' of time-varying objects (from which astronomers may query, search, and retrieve events based upon prescribed characteristic light curve behaviors) may prove to be one of the most significant contributions of citizen scientists to the LSST project -- i.e., the development of a major externally joined database component of the LSST science data collection.

Something like this approach was effectively used with the detection of the ``Peas" described above (Section 1.4.2.4) , i.e., once this class of objects was discovered, and determined to be interesting, a computer algorithm was developed to find them \citep{Cardamone2009}. By finding new classes of data, volunteers can make major contributions to science that would not be possible without their help.  The Galaxy Zoo Supernova project also uses a similar methodology \citep{Smith2011}.   During an observing run, the science team receives tens of thousands of possible supernova candidates, and    automatic algorithms are used to reduce this to a few hundred events a day that are likely supernovae.   With the help of citizen scientists, these candidate supernovae are visually checked and thus confirmed for follow-up observations in real-time.

In summary, as the data rates increase, and we become further dependent on automatic classification algorithms, citizen scientists can play a crucial role in reviewing subsets of the data 
and identifying anomalies.    The algorithms can then be adapted by the science teams to increase their success rate (based on the visual checks).
The two methodologies will need to work in tandem and the process will likely be iterative.

\section{Galaxy Zoo in the Context of other Citizen Science Projects}
Galaxy Zoo is certainly not the only citizen science project. As mentioned in the beginning of this chapter, one of the inspirations for Galaxy Zoo was the Stardust@Home project \citep{Westphal2006}. This was one of the few projects at the time where volunteers were asked to participate in the data \emph{analysis} of a project rather than the data \emph{collection} phase of a project. Many of the historically significant citizen science projects, such as the Audubon Society's Christmas Bird Count program (started in 1900) and the American Association of Variable Star Observers variable star observations project (starting in 1911), were based on data \emph{collection}. With the advent of the internet as a distribution system, citizen science projects could move into work on data \emph{analysis}. Here we make a distinction between distributed analysis projects and distributed computation projects such as SETI@Home\footnote{ \tt http://setiathome.berkeley.edu/index.php}, which utilizes the computational power of over a million idle computers belonging to volunteers to process radio data looking for signals that could indicate extra-terrestrial intelligence. Distributed analysis projects require the brain - or the ``wetware" - of the volunteer to be engaged, not just their computer. One of the earliest distributed data analysis projects (dating from 2001) was Clickworkers which asked the public to count the number of craters on maps of the Martian surface returned by the Mars Orbiter Camera (MOC). While the project was successful in recruiting sufficient volunteers to identify over 800,000 MOC craters \citep{Gulick2010}, there are few scientific results published as of yet on this body of work. Clickworkers has morphed into a project with a more game--like interface\footnote{ \tt http://beamartian.jpl.nasa.gov/welcome} where the public are currently asked to ``tag" surface  features on images from the Mars Rovers, Spirit and Opportunity. Another project with a game--like interface is FoldIt!\footnote{ \tt  http://fold.it/portal/} which asks the public to help solve protein folding ``puzzles". Both the new Clickworkers and the FoldIt! projects require the user to download an application. In the context of distributed data analysis projects, Galaxy Zoo (and its successor projects in the Zooniverse) is quite probably the largest both in terms of number of registered volunteers world--wide as well as number of peer--reviewed papers published based on data processed by volunteers.

There are many excellent citizen science projects in ecology, animal studies and other disciplines where the distributed nature of data \emph{collection} is critical to the success of the project. For example Cornell University's Lab of Ornithology FeederWatch Project\footnote{ \tt http://www.birds.cornell.edu/pfw/} asks volunteers to enter counts of bird species into an online form to track winter bird populations, or the CoCoRaHS Network (Community Collaborative Rain, Hail and Snow Network)\footnote{ \tt http://www.cocorahs.org/} has thousands of volunteers across all fifty of the United States who have installed sensors outside their homes and record amounts of rain, snow and hail. Thus, one potential trend in citizen science will then be projects that link the distributed data collection and distributed data analysis aspects of their work. 

\section{Conclusions}
In this chapter we have presented a brief overview of the Galaxy Zoo and Zooinverse projects. We gave a short discussion of the history and motivation for the original Galaxy Zoo, as well as the motivations to extend it to ``Galaxy Zoo 2", ``Hubble Zoo" and an entire ``Zooiniverse" of citizen science projects. We have described the highlights of the many scientific results that have already come from Galaxy Zoo. 

 We go on to discuss what makes a good citizen science project, and why we think Galaxy Zoo was so successful. We describe the importance of having a central portal, the Zooniverse, as a gateway to citizen science projects across multiple disciplines. We then consider likely future applications of community--based science in the coming data--rich era.
 Finally, to provide a context for the importance of the Galaxy Zoo project, we present a short description of various other citizen science projects and modalities.
 
\vspace*{0.3cm}

Galaxy Zoo and the many Zooniverse projects would not have been possible without the participation of now over 400,000 volunteers who have registered with the Zooinverse. The contributions of volunteers to Galaxy Zoo are individually acknowledged at {\tt http://www.galaxyzoo.org/Volunteers.aspx}; and volunteers who classified in Galaxy Zoo 2 (and wished to be acknowledged) are listed at {\tt http://zoo2.galaxyzoo.org/authors}.
The work described in this paper is funded
in part by The Leverhulme Trust (UK), the National Science Foundation and the National Aeronautics and Space
Administration (US).


\bibliographystyle{apj}

\end{document}